\documentclass[runningheads]{llncs}
\usepackage{graphicx}
\begin{document}
\title{Real-time FPGA Design for OMP Targeting 8K Image Reconstruction}
\author{Jiayao Xu, Chen Fu, Zhiqiang Zhang, Jinjia Zhou}
\authorrunning{ et al.}
\institute{Hosei University
\email{jiayao.xu.5k@stu.hosei.ac.jp}}

\authorrunning{Xu. et al.}
\maketitle              % typeset the header of the contribution
\begin{abstract}
During the past decade, implementing reconstruction algorithms on hardware has been at the center of much attention in the field of real-time reconstruction in Compressed Sensing (CS). Orthogonal Matching Pursuit (OMP) is the most widely used reconstruction algorithm on hardware implementation because OMP obtains good quality reconstruction results under a proper time cost. OMP includes Dot Product (DP) and Least Square Problem (LSP). These two parts have numerous division calculations and considerable vector-based multiplications, which limit the implementation of real-time reconstruction on hardware. In the theory of CS, besides the reconstruction algorithm, the choice of sensing matrix affects the quality of reconstruction. It also influences the reconstruction efficiency by affecting the hardware architecture. Thus, designing a real-time hardware architecture of OMP needs to take three factors into consideration. The choice of sensing matrix, the implementation of DP and LSP. In this paper, a sensing matrix, which is sparsity and contains zero vectors mainly, is adopted to optimize the OMP reconstruction to break the bottleneck of reconstruction efficiency. Based on the features of the chosen matrix, the DP and LSP are implemented by simple shift, add and comparing procedures. This work is implemented on the Xilinx Virtex UltraScale+ FPGA device. To reconstruct a digital signal with 1024 length under 0.25 sampling rate, the proposal method costs 0.818$\mu$s while the state-of-the-art costs 238$\mu$s. Thus, this work speedups the state-of-the-art method 290 times. This work costs 0.026s to reconstruct an 8K gray image, which achieves 30FPS real-time reconstruction.

\keywords{Compressed Sensing \and Reconstruction Algorithm  \and Orthogonal matching pursuit (OMP) \and Field Programmable Gate Array (FPGA) \and Real-time Reconstruction.}
\end{abstract}
\section{Introduction}
Nowadays, digital signals, such as images, connect with people's life closely. The traditional method to obtain the digital signals is based on the Nyquist-Shannon sampling theory. This theory indicates that to sample the digital signals, the digital sampling frequency must be at least twice the highest frequency of the original analog signal. This leads to numerous redundancy in the digital signals and requires considerable storage. Thus, compression is an essential procedure after sampling. With the development of the technology, this compression after sampling method can not match the requirement of the explosive growth of the digital signals.

To sample the digital signal, in the same time, compress it, the D Donoho, E Candes, T Tao and other scientists proposed a theory named Compressed Sensing (CS) [3] in 2006. Compressed Sensing guarantees that signals can be reconstructed effectively with few samples far less than Nyquist-Shannon theory requires. Different from the traditional method, the compression procedure of Compressed Sensing is just a simple linear projection while the reconstruct procedure is much complicated. Hence, the choice of the reconstruction algorithm is crucial to the reconstruction efficiency and quality. 

Orthogonal Matching Pursuit (OMP) [12] is one of the classic reconstruction algorithms in Compressed Sensing, which can achieve good quality results in a shorter time. Recently, implementing OMP on Field Programmable Gate Array (FPGA) to achieve real-time reconstruction has gained attention from researchers. The reconstruction procedure of OMP includes Dot Product (DP) and Least Square Problem (LSP). DP is applied by the vector-based multiplications mainly. LSP has division calculations and numerous vector-based multiplications. Because of the limitation of the hardware, division calculations demand a big overhead and need to be avoided. Thus, implementing LSP is the key difficulty for implementing OMP on hardware.

%\begin{figure}
%\includegraphics[width=\textwidth]{omp1.jpg}
%\caption{OMP reconstruction diagram} \label{fig1}
%\end{figure}

Many methods are proposed to substitute the LSP. In the relevant literature, there are two main approaches, the Cholesky decomposition method [2,8] and the QR decomposition [1,10,11]. The first OMP on FPGA framework was designed by Septimus and Steinberg [9] in 2010. They used the Cholesky decomposition to prevent division calculations. But, the Cholesky decomposition has high computation complexity, the QR decomposition was proposed to overcome this problem. Because QR decomposition contains square root calculation, Ge et al. [5] presented a Square-Root-Free QR Decomposition method to decline the reconstruction time. 

Besides the works illustrated above, to decrease the reconstruction time, other methods are adopted by the existing works. Li et al. [6] presented an approach using Gram–Schmidt Orthogonalization to substitute the LSP calculations in the iterations of reconstruction. In this method, the LSP is only be used once in the reconstruction procedure. Fardad et al. [4] proposed a deterministic matrix to replace the random matrix to decrease the reconstruction difficulty and save the hardware cost. 

Because of the complexity of the image signals, most of existing works are sparsity signal based, only several works [1,10,11] verified their works on natural images. However, none of them achieves real-time image reconstruction when the image size is bigger than 1080p.

Not only the reconstruction algorithm but also the sensing matrix is a key point that affects the reconstruction quality. The sensing matrix also influences the reconstruction efficiency by affecting the design of hardware architecture. Thus, choosing a proper sensing matrix is also a key point for hardware implementation.

To fulfill the goal of real-time image reconstruction, the main contributions of this paper include the following aspects.

1. In the proposed architecture, a sensing matrix obtained by multiplying the Hadamard matrix with the Fast Walsh-Hadamard transform matrix is adopted. This chosen sensing matrix is sparse and includes zero vectors mainly. And the reconstruction results by this sensing matrix has good quality. 

2. Based on the features of the chosen sensing matrix, the DP and LSP are applied by simple add, shift and comparing calculations to replace the complicated vector-based multiplications and division calculations.

3. Because the image signals are complicated, to guarantee the quality of the reconstructed results, the sparsity level of this architecture is not set as a certain value but in a range.

4. This proposed architecture achieves the real-time reconstruction of 8k 30FPS under 0.25 sampling rate.

\section{Background}
\subsection{Compressed Sensing (CS)}
Assuming the length of original image signal $X$ is $N*1$. Using a transform matrix marked as $\psi$ with the size of $N*N$, transforming the original signal to the transform domain to gain the sparse signal called $\theta$ with the size of $N*1$. Sparse means, in the signal, there are only a few elements are non-zero elements. The sparsity level is marked as $k$. Projecting the original signal on the measurement matrix $\phi$, then the compressed signal $Y$ called measurements is obtained. The size of measurement matrix is $M*N$, the size of signal $Y$ is $M*1$. The relationship between $k,M,N$ is $K<<M<N$. The $M/N$ represents the sampling rate. Multiplying the measurement matrix with the transform matrix is the sensing matrix $A$. The theory of CS is as Equation.1 shows. 
\begin{equation} 
Y = \phi_{M*N}X_{N*1} = \phi_{M*N}\psi_{N*N}\theta_{N*1} = A_{M*N}\theta_{N*1}  
\end{equation}

Compressed Sensing has two main principles. One is the sparsity of the original signals. This means the signal is compressible. The other one is incoherence. The measurement matrix with the transform matrix should have incoherence. The incoherence ensures the possibility of reconstructing the original. In other words, the sensing matrix, which is gained by multiplying the measurement matrix with the transform matrix, determines the quality of the reconstruction results. Thus, choosing proper sensing matrix is important.

\subsection{Orthogonal Matching Pursuit (OMP)}
Orthogonal Matching Pursuit (OMP) is a reconstruction algorithm that belongs to the greedy algorithm. To reconstruct the original image signal, OMP first uses Dot Product to find the index of the column vector of the sensing matrix with the highest contribution, then puts this index into the index set. After that, the column vectors in the index set are used to approximate the original signal and update the residual. Both the procedure of approximating the original signal and the procedure of updating the residual include the Least Square Problem. The procedure of OMP reconstruction is as follows:

a. Initialize the residual $r_0 = Y$, the iteration counter $t=1$, the index set $S_0 = \emptyset$. 

b. Find the index of the highest contribution column vector of the sensing matrix. $Index_t$ = $argmax|A^T*r_{t-1}|$.

c. Update the index set $S_t = S_{t-1} \cup Index_t$.

d. Use Least Square Problem to approximate the original signal.
$\hat{\theta} = (A_S^T*A_S)^{-1}A_S^T*Y$

e. Judge if satisfy the quit condition or not. If yes, go to g. If not, go to f.

f. Update residual. $r_t = Y - A_S*\hat{\theta}$, then go back to b.

g. Inverse the approximate signal to obtain the reconstructed image signal. $\hat{X} = \psi_{N*N}*\hat{\theta}$.

OMP is an iteration-based reconstruction algorithm. If the reconstruction procedure satisfies the quit condition, the approximation signal will multiply with the transform matrix to obtain the reconstructed image signal. The iteration time is usually regarded as the sparsity level of the signal in the transform domain. But the sparsity is unknown for practical image reconstruction. If the iteration time is too large, the reconstruction time will be considerable. If the iteration time is too less, the quality of the result will be worse. Thus, the strategy of the quit condition is crucial in the design.

\section{Proposals}
\subsection{Chosen Matrices and Features of Sensing matrix}
In this work, the Hadamard matrix is adopted as the measurement matrix, the Fast Walsh-Hadamard transform matrix is chosen as the transform matrix. The sensing matrix $A$, which is used in the reconstruction procedure, is as the Equation.2 and Fig.1 shows.

\begin{equation}
A\left(i,j\right)=\left\{
    \begin{array}{l}
        2*V,i=j=1\\
        V,  i=j \cup i=1 \cap j\neq1\\
        0,others
    \end{array}
\right.
\end{equation}

\begin{figure}
\includegraphics[width=\textwidth]{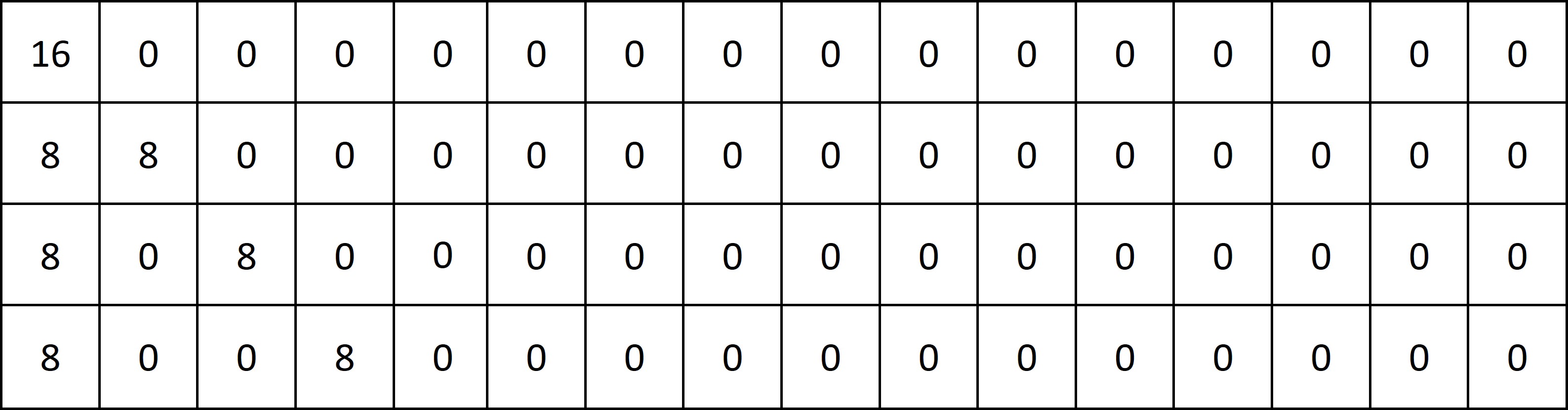}
\caption{An example of Sensing Matrix ($M = 4$, $N = 16$, Sampling rate is 0.25)} 
\end{figure}

In the Equation.2, the value of $V$ changes with the different choices of M and N. Especially, the range of $i$ is $0<i\leq M$, the range of $j$ is $0< j\leq N$. It is easy to find that when $j>M$, $A(i,j)$ is zero. That is, if the vector index is bigger than $M$, the column vectors are all zero vectors. Besides, the non-zero column vectors of the chosen sensing matrix are sparse vectors, which contain zero elements mainly. Based on these features, the Dot Product and Least Square Problem are optimized further to reduce the time cost to achieve the aim of real-time reconstruction.

\subsection{Optimized Dot Product}
This subsection introduces the optimized Dot Product implementation. The Dot Product is used to find the index of the highest contribution column vector of the sensing matrix. The original calculation is illustrated in Equation.3. 
\begin{equation} 
Index = argmax|A^T*r|
\end{equation}
Because the column vectors with index bigger than $M$ are all zero vectors, the vector-based multiplication results of these vectors are zero. Hence, the Dot Product in Equation.3 is optimized as Equation.4 shows.
\begin{equation} 
result_{DP} = A^T_{j,j=1..M}*r
\end{equation}
In addition, the column vectors from 1 to $M$ are sparse vectors, the calculation in Equation.3 can be revised further. Deleting the calculations of zero elements, the Dot Product  calculation is as Equation.5 shows.

\begin{equation} 
result_{DP}(j)= V * \left\{
    \begin{array}{l}
        2*y_1 + y_2 + ... + y_M, j=1\\
        y_j,  1 < j \leq M
    \end{array}
\right.
\end{equation}

After obtaining the result of the Dot Product, this result will be sort to find the index of the highest contribution column vector of the sensing matrix. This sort procedure is implemented by 2-to-1 comparison calculation. As Equation.5 presents, there is a common divisor in the result of Dot Product. The multiplication with this common divisor, which is represented as $V$ in Equation.5, can be deleted to save both time and hardware cost. 

\begin{equation} 
Index = argmax |result_{DP}(j)| 
\end{equation}
\begin{equation} 
result_{DP}(j)= \left\{
    \begin{array}{l}
        0, j=1\\
        y_j,  1 < j \leq M
    \end{array}
\right.
\end{equation}

Because in the first iteration, the residual is equals to the measurements, and the values of measurements are positive numbers, the $result_{DP}(1)$ must be the biggest value. That is, the $Index$ of the first iteration must be 1. In OMP, all of the column vectors only can be chosen once. Based on these conditions, in the first iteration, the procedure in Equation.3 is omitted, the result of $Index$ in the first iteration is set as 1 and the $result_{DP}(1)$ in other iterations is set as 0. The implemented procedure of finding the index of the highest contribution column vector is as Equation.6, Equation.7 present. 

\begin{figure}
\includegraphics[width=\textwidth]{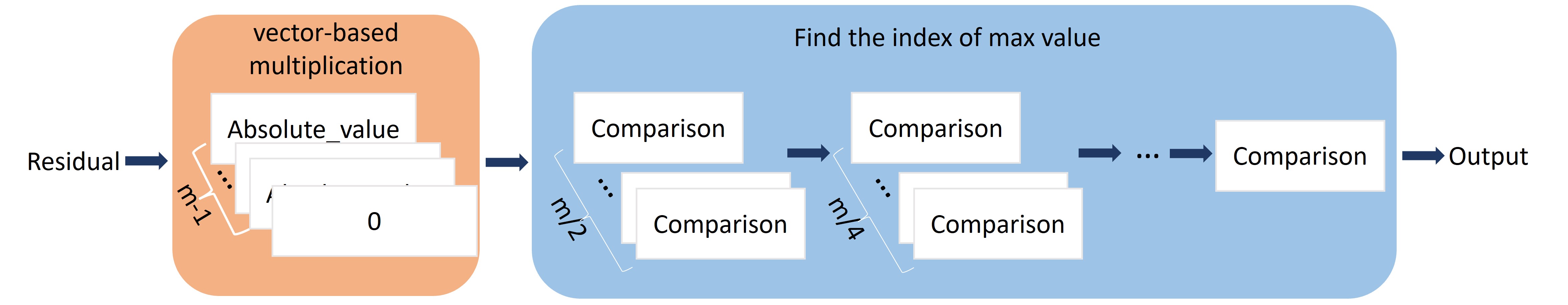}
\caption{Finding the index of the highest contribution column vector} 
\end{figure}

Fig.2 shows the diagram of the implemented procedure when the iteration index is bigger than 1. First, the first element of the residual is set as 0, other elements are set as its absolute value. Then using the 2 to 1 comparison to find the biggest value. Last, outputting the index of the biggest value.

\subsection{Optimized Least Square Problem}
This subsection introduces the optimized Least Square Problem (LSP). The LSP is the key point at both residual update in the iteration and the original signal approximation after quitting the iteration. The calculation of LSP is as Equation.8 shows. 

\begin{equation} 
\hat{\theta} = (A_S^T*A_S)^{-1}A_S^T*Y
\end{equation}

As mentioned in Section 3.1, the $A_S$ contains sparse column vectors. Hence, in the vector-based multiplication of $A_S^T*A_S$, only several situations, which are illustrated in Equation.9, get non-zero results. 

\begin{equation} 
Result_{Non-zero} = \left\{
    \begin{array}{l}
        A_S^T(1)*A_S(j), \\
        A_S^T(j)*A_S(j),  \\
        A_S^T(j)*A_S(1), 
    \end{array}
\right.
 1 < j \leq M
\end{equation}

Marking the $A_S^T*A_S$ as $Result_{matrix}$, the Equation.10 shows the equation of $Result_{matrix}$.

%\begin{figure}
%\includegraphics[width=\textwidth]{non-zero results.jpg}
%\caption{Situations of non-zero results ($M = 4$, $N = 16$, Sampling rate is 0.25)} 
%\end{figure}

\begin{equation} 
Result_{matrix}(i,j) = \left\{
    \begin{array}{l}
        P,  i=j=1\\
        Q,  i=1 \cap j\neq 1\\
        Q,  j=1 \cap i\neq 1\\
        Q,  i = j \cap i=j\neq 1
    \end{array}
\right.
\end{equation}

The range of $i$ is $ 1 \leq i \leq index_{iteration}$, the range of $j$ is $1 \leq j \leq index_{iteration}$. The $index_{iteration}$ represents the iteration index. The value of $P$ and $Q$ depends on the iteration index.

Because the value of matrix $Result_{matrix}$ is certain in each iteration, the value of $Result_{matrix}^{-1}$ is certain too. Hence, the complicated inverse calculation is substituted as using a Look Up Table (LUT) to find the value of $Result_{matrix}^{-1}$ of each iteration. But the inverse value leads to fractional numbers which is hard to implement in hardware, shifting the input is indispensable. 

The $A_S^T*Y$ in Equation.8 is still a time cost procedure. As mentioned in Section 2.2, the index set $S$ unions the max index of the current iteration with the indexes stored in the previous iterations. The $A_S^T*Y$ also can be replaced by combining the result of previous iteration with the result of the current iteration. So only the new result of $A_S^T*Y$ is calculated in each iteration. Besides, the column vectors of $A_S^T$ are sparsity, so the vector-based multiplication can also be substituted by the simple shift and add calculations. 

\begin{figure}
\includegraphics[width=\textwidth]{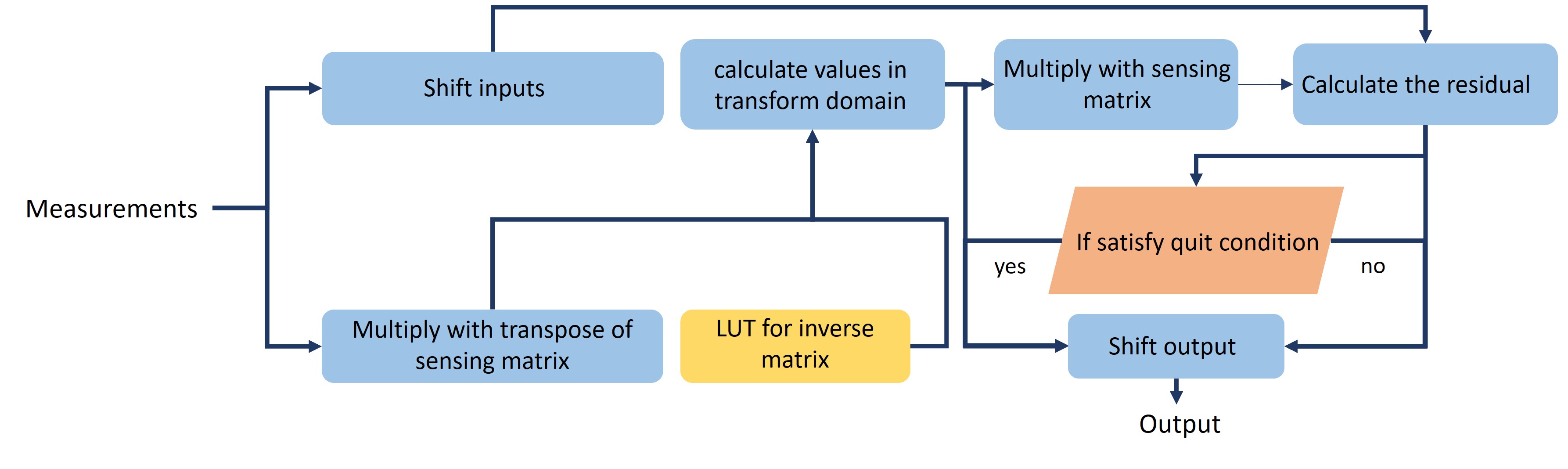}
\caption{The implementation of the Least Square Problem} 
\end{figure}

The implementation details of the Least Square Problem is as Fig.3 shows. First, multiply measurements with the transpose of sensing matrix. Then multiply it with the inverse matrix, which is stored as LUTs, to obtain the approximate signals in transform domain. Next, multiply the approximate signals in transform domain with the sensing matrix. After that, use the shift measurements to update the residual. Judge if meets the quit condition or not. If yes, output the shift value of the approximate signals in transform domain. If not, output the shift values of updated residual.

\subsection{Overall Framework}

\begin{figure}
\includegraphics[width=\textwidth]{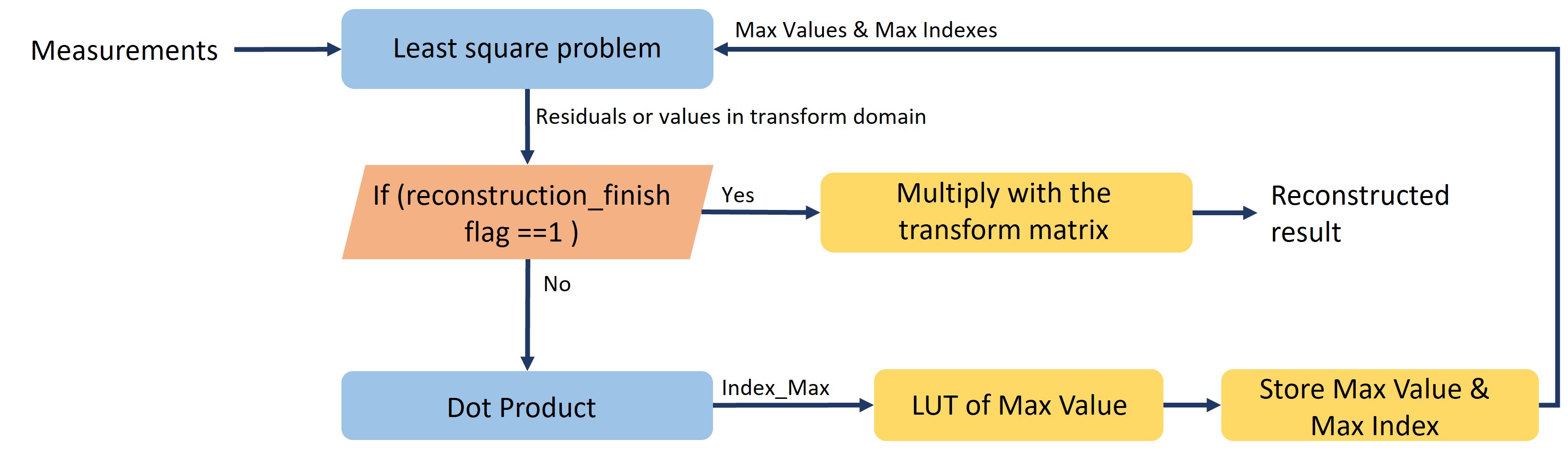}
\caption{The overall architecture} 
\end{figure}

Fig.4 shows the overall architecture of this work. As illustrated in Section 3.2, the first iteration of the reconstruction omits the Dot Product procedure. The result of the Dot Product in the first iteration is defaulted as 1. So the reconstruction procedure of this work starts from LSP to update the residual. In the LSP, if the quit condition is satisfied, LSP will output a flag named $reconstruction\_finish$ and the shift of the approximate signals in transform domain. If the $reconstruction\_finish$ is not equals to 1, the output residual of LSP will put into the Dot Product to find the index of the highest contribution column vector of the sensing matrix. As mentioned in Section 3.3, there are calculations duplicate in every iteration. Thus, the max value of residual and max index of column vector are stored in every iteration. These parameters are inputs of LSP. This loop will be quit until the  $reconstruction\_finish$ is 1. Then multiplying the output of the LSP with the transform matrix to obtain the reconstructed result.

\section{Experimental Results}
\subsection{Design Details}
\subsubsection{Quality Comparison with Gaussian}
Most of the existing works adopt the Gaussian matrix as the measurement matrix to verify their proposals. The Gaussian matrix is the classic random matrix of Compressed Sensing. But it consists of floating values, which is difficult to implement on the hardware. To compare with the matrix chosen by the existing work, 200 images with the size of 320*480 are chosen from the database BSD 500 [7] to do the experimental test. The transform matrix to test the Gaussian matrix is Discrete Cosine Transform (DCT). The MATLAB is used to do this comparison test. As Table.1 shows, the result quality of matrices chosen by this work outperforms that of the Gaussian matrix. 

\begin{table}
\caption{Result quality comparison}
\centering
\begin{tabular}{|l|l|l|l|}
\hline
Matrix name & Sampling rate & PSNR & SSIM \\
\hline
Gaussian with DCT &  0.25 & 19.86 & 0.312\\
Matrices of this work &  0.25 & {\bfseries 26.30} & {\bfseries 0.640}\\
Gaussian with DCT &  0.5 & 23.85 & 0.550\\ 
Matrices of this work &  0.5 & {\bfseries 29.48} & {\bfseries 0.818}\\
Gaussian with DCT &  0.75 & 27.02 & 0.726\\
Matrices of this work &  0.75 & {\bfseries 32.37} & {\bfseries 0.908}\\
\hline
\end{tabular}
\end{table}

\subsubsection{Fixed Point}
In Section 3.3, the inverse of the $Result_{matrix}$ is stored in the LUTs. The inversion procedure leads to fractional numbers. To implement the fractional numbers on hardware, shifting data in LSP is adopted. Due to the limitation of the hardware, the fractional numbers only can be set as fixed point. Fixed point decreases the quality of the reconstruction results. To decline the influences of the fixed point on the result, the test of choosing the proper fixed point is implemented in MATLAB using the same test images as Quality Comparison with Gaussian uses. 

\begin{table}
\caption{Test result of fixed point (Sampling rate is 0.25) }
\centering
\begin{tabular}{|l|l|l|}
\hline
Fixed point & PSNR & PSNR increase\\
\hline
8 &  25.23 & -\\
9 &  26.06 & 0.836\\
10 & 26.19 & 0.124\\ 
11 & 26.26 & 0.076\\
12 & 26.28 & 0.013\\
\hline
\end{tabular}
\end{table}

The Table.2 shows the test result of the fixed point under the sampling rate of 0.25. The PSNR increase is calculated by using PSNR of the current line to diminish the PSNR of the upper line. Because the PSNR increase of fixed point 12 is lower than 0.05, the fixed point of this work is set as 11.

\subsubsection{Quit Condition and Sparsity}
As mentioned in Section 2.2, the sparsity of the natural images is unknown. If the sparsity is too large, the iteration time will be high then the reconstruction time will be numerous. If the sparsity is too small, the quality of the results will be worse. To balance this trade-off, in the proposed architecture, the sparsity is set as 8, which is half of the $M$. If the iteration index equals 8, the iteration will be quit. Besides, if the residual in the iteration equals 0, the iteration will be quit too. 

\subsubsection{Design parameters} 
To lay the foundation on the future works, the experimental images are separated into 8*8 blocks to do the compression. The reconstruction architecture is implemented by $N = 64$, $M = 16$, $k = 8$. 

\subsection{Design Performance}

\subsubsection{Design Verification Environment}
Verilog is adopted to describe the proposed architecture. The proposed architecture is synthesised and implemented by the Xilinx Vivado 2020.2 on Virtex UltraScale+ FPGA. 

\subsubsection{Hardware Cost and Reconstruction Time Comparison}
In the existing works, the signal length is set as 1024. To compare the hardware cost and reconstruction time with the existing works, this paper initializes 16 reconstruction blocks with the length of 64. The comparison details are as Table.3 presents. 

\begin{table}
\caption{Comparison this work with existing works}
\centering
\begin{tabular}{|l|l|l|l|l|}
\hline
& This work & [1] & [4] & [5]\\
\hline
FPGA &  Virtex UltraScale+ & Virtex-6 & Virtex-6 & Kintex-7 \\
Signal size & 1024 & 1024 & 1024 & 1024\\
Number of measurement & 256 & 256 & 248 & 256\\ 
Sparsity & (up to) 128 & 36 & 36 & 36\\
Sparsity fixed & No & Yes & Yes & Yes\\
Frequency (MHz) & 133.33 & 100 & 123 & 210\\
Reconstruction time ($\mu$s) & 0.818 & 622 & 333 & 238\\
Occupied Slices & 47624 & 32010 & 853 & 28443\\
LUT & 266542 & - & - & 59414\\
FF & 64713 & - & - & -\\
DSP & 80 & 261 & 7 & 523\\
BRAM & 0 & 258 & 4 & 386\\
\hline
\end{tabular}
\end{table}

Comparing with the existing 3 works, only this work is sparsity unfixed, which makes the proposed design more flexible. Because the sensing matrix is embedded in the reconstruction procedure, this work costs 0 BRAM. The most notable point is that the reconstruction time of this work surpasses the existing works obviously. Reconstructing a signal with length of 1024 only needs 0.818us. 

\begin{table}
\caption{Reconstruction time under different image sizes (Sampling rate is 0.25)}
\centering
\begin{tabular}{|l|l|}
\hline
Resolution & Reconstruction time (s)\\
\hline
1080p (1920x1080) &  0.0017 \\
4K (3840×2160) &  0.0066 \\
8K (7680×4320) & 0.0265 \\ 
\hline
\end{tabular}
\end{table}

Table.4 shows the reconstruction time under different image sizes. The proposed architecture can afford 1080p (1920x1080)$\times$120 FPS, 4K (3840×2160)$\times$120 FPS and 8K (7680×4320)$\times$30 FPS real-time reconstruction. 

Fig.5 illustrates an example of reconstructed image result under sampling rate of 0.25. It is well known that if an image contains more backgrounds, it has more low-frequency information at the transform domain. On the other hand, if an image contains more details, it has more high-frequency information at the transform domain. Fig.2 shows that if the index to the column vector of the sensing matrix is bigger than $M$, the column vectors are all zero vectors. In other words, this sensing matrix keeps more low-frequency information at the transform domain. To compare the reconstruction quality between high-frequency image and low-frequency image, there are two block images of the reconstructed image are selected. In Fig.5, the left block image contains more backgrounds while the right block image contains more details. The reconstruction quality of the left image outperforms than right image. Thus, the proposed hardware is more friendly to images with low-frequency information mainly.

\section{Conclusion}
This paper introduces a hardware implementation of OMP named Real-time FPGA Design for OMP Targeting 8K Image Reconstruction. In the implementation of OMP on hardware, there are three factors that infect the efficiency and quality of the results. The choice of sensing matrix, the design of Dot Product (DP), and the design of Least Square Problem (LSP). In this work, a sparse sensing matrix, which contains zero vectors mainly, is obtained by multiplying the Hadamard matrix with the Fast Walsh-Hadamard transform matrix. The features of this sensing matrix make that optimizing the DP and LSP is possible. The complicated divisions and vector-based multiplications of DP and LSP are replaced by the simple shift, add, and comparing calculations. These optimizations make achieving the goal of 8K real-time reconstruction possible. The result shows that to reconstruct an 8K gray image, the proposed architecture costs only 0.026s with reasonable quality. Comparing with the State-of-the-art method, the proposed architecture speedups the reconstruction efficiency with 290 times. And the proposed architecture meets the requirement of reconstructing 8K$\times$30 FPS real-time. 

\begin{figure}
\includegraphics[width=\textwidth]{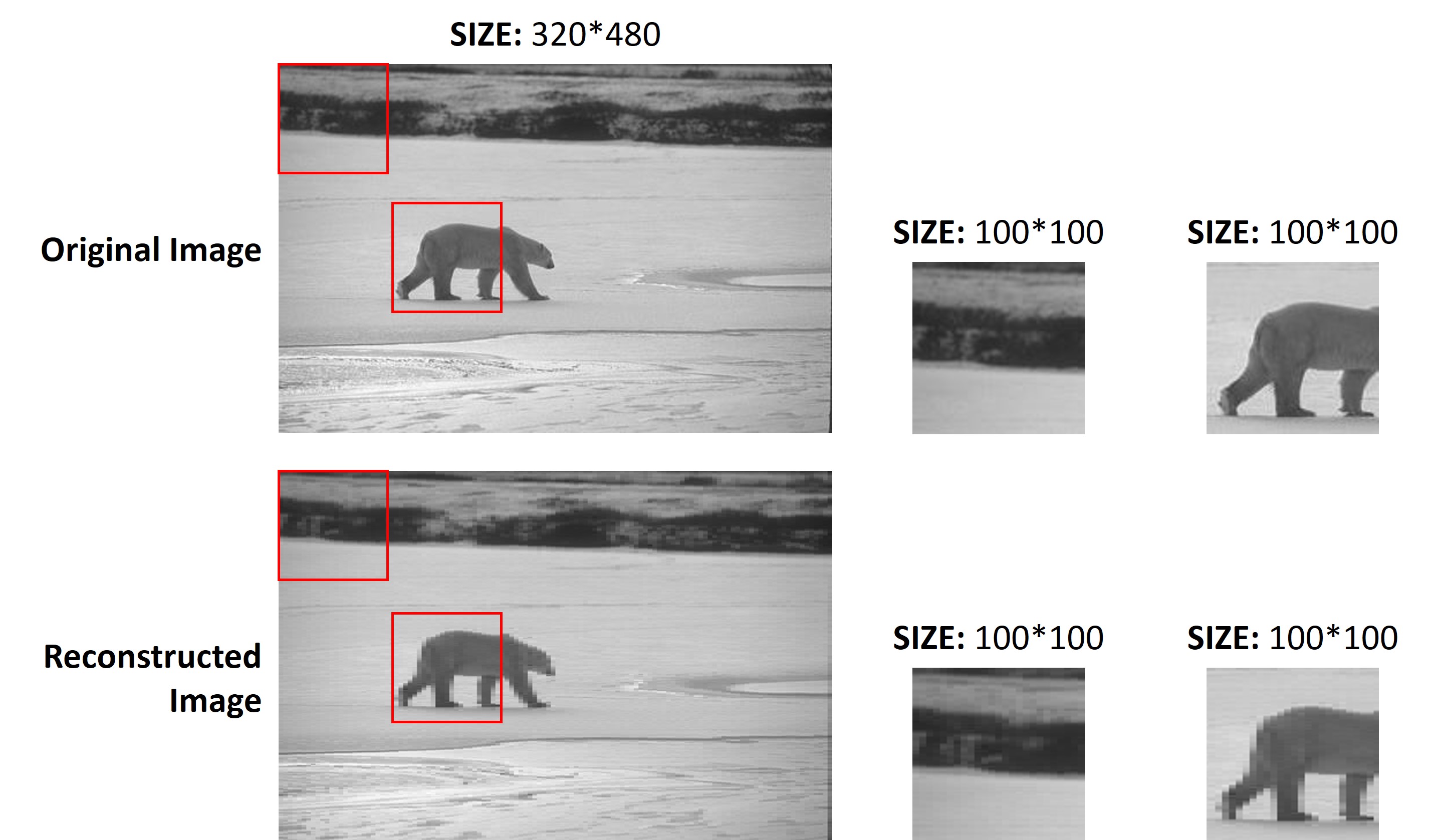} 
\caption{An example of reconstructed image result (Sampling rate is 0.25). The PSNR of this reconstructed image is 32.3647, the SSIM of this reconstructed image is 0.7334. Two block images of the reconstructed image with the size of 100*100 are selected. Block image at left contains backgrounds mainly, block image at right contains details mainly. } 
\end{figure}

\section{Reference}
1. Bai, L., Maechler, P., Muehlberghuber, M., Kaeslin, H.: High-speed compressed
sensing reconstruction on fpga using omp and amp. In: 2012 19th IEEE International Conference on Electronics, Circuits, and Systems (ICECS 2012). pp. 53{56.
IEEE (2012)

2. Blache, P., Rabah, H., Amira, A.: High level prototyping and fpga implementation
of the orthogonal matching pursuit algorithm. In: 2012 11th International Conference on Information Science, Signal Processing and their Applications (ISSPA).
pp. 1336{1340. IEEE (2012)

3. Donoho, D.L.: Compressed sensing. IEEE Transactions on information theory
52(4), 1289{1306 (2006)

4. Fardad, M., Sayedi, S.M., Yazdian, E.: A low-complexity hardware for deterministic
compressive sensing reconstruction. IEEE Transactions on Circuits and Systems I:
Regular Papers 65(10), 3349{3361 (2018)

5. Ge, X., Yang, F., Zhu, H., Zeng, X., Zhou, D.: An efficient fpga implementation of
orthogonal matching pursuit with square-root-free qr decomposition. IEEE Transactions on Very Large Scale Integration (VLSI) Systems 27(3), 611{623 (2018)

6. Li, J., Chow, P., Peng, Y., Jiang, T.: Fpga implementation of an improved omp
for compressive sensing reconstruction. IEEE Transactions on Very Large Scale
Integration (VLSI) Systems 29(2), 259{272 (2020)

7. Martin, D., Fowlkes, C., Tal, D., Malik, J.: A database of human segmented natural images and its application to evaluating segmentation algorithms and measuring ecological statistics. In: Proceedings Eighth IEEE International Conference on
Computer Vision. ICCV 2001. vol. 2, pp. 416{423. IEEE (2001)

8. Rabah, H., Amira, A., Mohanty, B.K., Almaadeed, S., Meher, P.K.: Fpga implementation of orthogonal matching pursuit for compressive sensing reconstruction.
IEEE Transactions on very large scale integration (VLSI) Systems 23(10), 2209{
2220 (2014)

9. Septimus, A., Steinberg, R.: Compressive sampling hardware reconstruction. In:
Proceedings of 2010 IEEE International Symposium on Circuits and Systems. pp.
3316{3319. IEEE (2010)

10. Stanislaus, J.L., Mohsenin, T.: High performance compressive sensing reconstruction hardware with qrd process. In: 2012 IEEE international symposium on circuits
and systems (ISCAS). pp. 29{32. IEEE (2012)

11. Stanislaus, J.L., Mohsenin, T.: Low-complexity fpga implementation of compressive sensing reconstruction. In: 2013 International conference on computing, networking and communications (ICNC). pp. 671{675. IEEE (2013)

12. Tropp, J.A., Gilbert, A.C.: Signal recovery from random measurements via orthogonal matching pursuit. IEEE Transactions on information theory 53(12), 4655{
4666 (2007)

%\bibliographystyle{splncs04}
%\bibliography{MMM}
\end{document}